\newcommand{\nn}{\nonumber}
\newcommand{\bsub}{\begin{subequations}}
\newcommand{\esub}{\end{subequations}}
\newcommand{\beq}{\begin{equation}}
\newcommand{\eeq}{\end{equation}}
\newcommand{\beqa}{\begin{eqnarray}}
\newcommand{\eeqa}{\end{eqnarray}}
\newcommand{\beql}{\begin{subequations}\begin{eqnarray}}
\newcommand{\eeql}{\end{eqnarray}\end{subequations}}
\documentclass[pra,aps,amsmath, twocolumn, nofootinbib]{revtex4-1}
\usepackage{layout}
\usepackage[english]{babel}
\usepackage{amsmath} 
\usepackage{graphicx}
\usepackage{float}   
\usepackage{verbatim}  
\usepackage{braket}
\usepackage{color}
\usepackage{cases}

\begin{document}
\title{Critical behaviour of coherence and correlation of counterpropagating twin beams}
\author{T.~Corti$^1$, E.~Brambilla$^{1}$, and A.~Gatti$^{1,2}$}
\affiliation{$^1$ Dipartimento di Scienza e Alta Tecnologia dell'Universit\`a dell'Insubria, Via Valleggio 11 Como, Italy, \\
$^2$ Istituto di Fotonica e Nanotecnologie del CNR, Piazza Leonardo da Vinci 32, Milano, Italy}
\today
\begin{abstract}
This work analyses the temporal coherence and correlation of twin beams generated in a quasi-phase matched nonlinear crystal in a counterpropagating configuration, ranging from the low-gain regime, where counterpropagating photon pairs are generated spontaneously, to the regime of stimulated pair production, close to the MOPO (Mirrorless Optical Parametric Oscillator) threshold. Here we show a critical divergence of the correlation time and slowing down of quantum fluctuations originating from the feedback mechanism responsible of the MOPO threshold.
\end{abstract}
\maketitle

\section{Introduction}
\label{sec:intro}
\begin{figure}[!htbp]
     \includegraphics[width=0.5\textwidth]{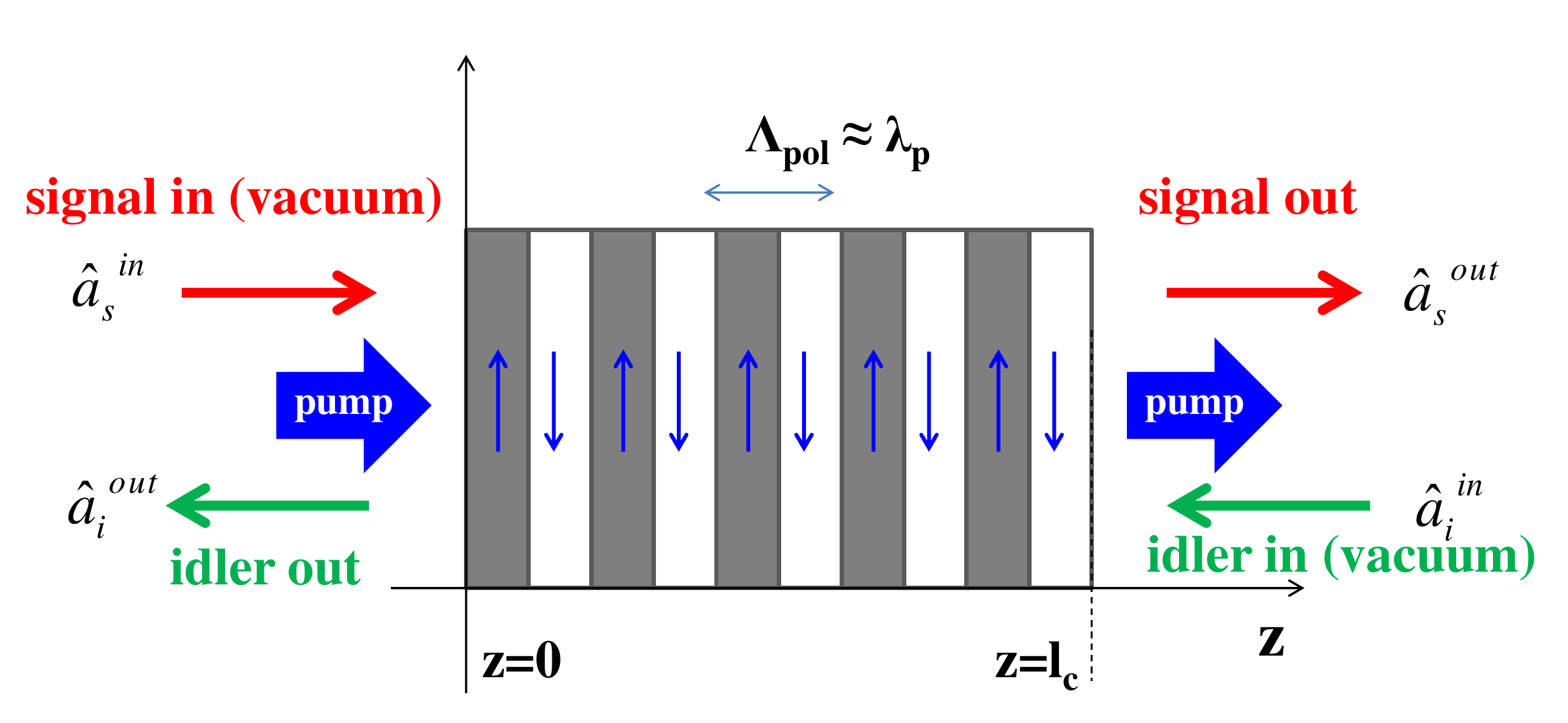}
           \caption{(Color online) Scheme of the counterpropagating parametric down-conversion based on a periodically poled crystal of length $l_c$ and poling period $\Lambda_{pol}\approx \lambda_p$. Quasi-phase matching determines the propagation of the idler field opposite to the pump and signal propagation directions (see text).}\label{fig:pp_configuration}
\end{figure}

Parametric down-conversion in a quasi-phasematched $\chi^{(2)}$ crystal with a poling period on the order of the pump field wavelength allows the generation of counterpropagating twin beams (Fig.\,\ref{fig:pp_configuration}). 
This occurs because in each elementary down-conversion process the momentum imparted by the nonlinear grating is sufficiently large to almost compensate for the pump photon momentum and, as a result, the emitted twin photons propagate along opposite directions in order to satisfy the momentum conservation. A unique feature of this counterpropagating geometry is the presence of distributed feedback which leads to a transition towards coherent oscillations when the pump intensity exceeds a given threshold value. This concept was proposed theoretically many years ago \cite{Harris}, but only recently experimental evidence has been achieved \cite{Can} due to the technical difficulties involved in the fabrication of periodically poled crystal with the required submicrometric poling period \cite{Busacca,Can2}. Above threshold the system can in principle be exploited as a source of coherent and tunable radiation, thereby the name mirrorless optical parametric oscillators (MOPO) \cite{Can}. The unique property of the MOPO is the ultra-narrow spectral bandwidth characterizing the backward propagating field, which can be smaller than that of the pump by about two orders of magnitude \cite{Can,Can2}. The unusual properties of temporal coherence of the MOPO radiation above threshold have been studied in \cite{Strom,Strom2,Montes14}. The three wave-mixing interaction with counterpropagating fields has been investigated also in different contexts \cite{Mu,Gallo,Longhi,Khurgin,Harris2011}. An overview can be found in \cite{Pasisk}.

In the regime of purely spontaneous down-conversion, the counter-propagating configuration offers the unique opportunity of generating highly monochromatic photon pairs in an almost separable state, so that it represents a promising source of  heralded single photons with a high degree of purity \cite{Christ}. In a parallel and complementary work \cite{Gatti} we made a systematic investigation of the degree of entanglement of counterpropagating twin photons generated in the purely spontaneous regime. 

In this paper we investigate the coherence and correlation properties of the twin beams generated in the MOPO below threshold, pumped by a stationary monochromatic field. Based on a model describing parametric down-conversion (PDC) within the undepleted pump approximation we analyze the transition from the regime far from threshold, where purely spontaneous down-conversion is the main source of twin photon pairs, up to a regime close to threshold where the combined effect of stimulated PDC and distributed feedback affects dramatically the properties of the light source. We provide an intuitive picture explaining the transition between these two regimes and illustrate the divergence of the correlation time and the critical slowing down phenomenon of temporal fluctuations occurring when approaching the transition towards coherent oscillations.

The paper is organized as follows: Sec. II introduces a general quantum model, describing the coupled propagation equations of pump, signal and idler field operators in the counterpropagating geometry. Sec III illustrates its classical counterpart and review the main features of the classical MOPO description in the the CW pump approximation. Sec. IV derives a linearized quantum model, equivalent to the one introduced in Ref.\,\cite{Suhara}, valid below the MOPO threshold in the CW pump approximation. With this model, in sec V, we derive the general expression for the correlation and coherence functions. In the last part of the paper we provide a detailed analysis of their behaviour both in the frequency domain and in time domain, illustrating what distinguishes the regime of single photon pair production (Sec.VI) from the regime close to the MOPO threshold (Sec.VII). An intuitive explanation of the transition between these two regimes is given in Sec.VIII.
\section{The general quantum model}
\label{sec:model}
Periodic poling in ferro-electric $\chi^{(2)}$ materials such as KTP or LiNbO3 allows to manipulate the phase-matching conditions for three-wave mixing interaction to a high extent. In the counter-propagating configuration, the poling period $\Lambda$ must be on the order of the pump field wavelength $\lambda_p$ (see Fig.\,\ref{fig:pp_configuration}). This allows to satisfy quasi-phase matching conditions at first order with one of the two down-converted waves, say the idler, propagating in the opposite direction with respect to the incident pump. Denoting with $\omega_p$ the pump central frequency, this occurs for those frequencies $\omega_s$ and $\omega_i=\omega_p-\omega_s$ of the signal and idler waves for which the corresponding wave numbers $k_j=\frac{\omega_j \, n_j(\omega_j)}{c}$ satisfy the condition
\beq
k_s-k_i-k_p+k_G=0,
\label{eq:pm}
\eeq
where $k_G=2\pi/\Lambda$ is the first order $k$-vector.

We restrict to a purely temporal description: we consider only collinear propagation, either assuming that a small angular bandwidth is collected and the process is characterized by a single spatial mode operation, or because of a waveguiding configuration. The pump and signal fields propagate along the $+z$ direction, while the idler propagates along the $-z$ direction (Fig.\,\ref{fig:pp_configuration}).

It is convenient to introduce the positive frequency parts of field operators for the three wavepackets as:
\bsub
\beqa
\hat{A}_s(\Omega,z)&=&e^{+i k_s(\Omega)z}\hat{a}_s(\Omega,z)\\
\hat{A}_i(\Omega,z)&=&e^{-i k_i(\Omega)z}\hat{a}_i(\Omega,z)\\
\hat{A}_p(\Omega,z)&=&e^{+i k_p(\Omega)z}\hat{a}_p(\Omega,z), \label{eq:lin_prop_p}
\eeqa
\label{eq:lin_prop}
\esub
where capital $\Omega$ denotes the frequency offset from the respective central frequencies $\omega_s$, $\omega_i$ and $\omega_p=\omega_s+\omega_i$, which satisfy the quasi-phasematching condition (\ref{eq:pm}). The phase factors $e^{\pm k_j(\Omega)z}$ in (\ref{eq:lin_prop}) account for linear propagation, $k_j(\Omega)$ denoting the wavenumber of the field $j$ at frequency $\omega_j+\Omega$ ($j=s,i,p$). The lower case operators $\hat{a}_j$ are thus slowly varying with respect to the original field operators $\hat{A}_j$ since they evolve only because of the nonlinear interaction.

Considering the simplest case of a periodic poling of the crystal with identical layers of alternate orientation, the effective nonlinear susceptibility $\chi(z)$ is a periodic function which can be expressed as a Fourier series of the form
\beq
\chi(z)=\sum_n \chi_n e^{i \frac{2\pi n}{\Lambda_{pol}} z},\;\;\;\;n=\pm 1,\pm 3,\pm 5,\cdots
\eeq
with the Fourier coefficients $\chi_n$ scaling as $1/n$.  Retaining only the lowest order terms, $n=+1$ for the signal and idler field and $n=-1$ for the pump, it can be shown that the operators $\hat{a}_j$ satisfy the following propagation equations:
\bsub
\begin{align}
&\frac{\partial}{\partial z} \hat{a}_s(z,\Omega)=+\chi\int d\Omega' \hat{a}_p(z,\Omega+\Omega')\hat{a}_i^{\dagger}(z,\Omega')e^{-i \mathcal{D}(\Omega,\Omega')z}\\
&\frac{\partial}{\partial z} \hat{a}_i(z,\Omega)=-\chi\int d\Omega' \hat{a}_p(z,\Omega+\Omega')\hat{a}_s^{\dagger}(z,\Omega')e^{-i \mathcal{D}(\Omega',\Omega)z}\\
&\frac{\partial}{\partial z} \hat{a}_p(z,\Omega)=-\chi\int d\Omega' \hat{a}_s(z,\Omega')\hat{a}_i(z,\Omega-\Omega')e^{i \mathcal{D}(\Omega',\Omega-\Omega')z}
\end{align}
\label{eq:prop_eq}
\esub
where the coupling costant $\chi\propto\chi_1=\chi_{-1}$ and
\beq
\mathcal{D}(\Omega,\Omega')=k_s(\Omega)-k_i(\Omega')-k_p(\Omega+\Omega')+\frac{2\pi}{\Lambda}
\eeq
is the effective phase mismatch, which rules the efficiency of each elementary down-conversion process, where a signal and an idler photon of frequencies $\omega_s+\Omega$, $\omega_i+\Omega'$ are generated out of a pump photon of frequency $\omega_p+\Omega+\Omega'$. It expresses the momentum balance including also the momentum $k_G=\frac{2\pi}{\Lambda}$ of the reciprocal lattice of the polarization inversion.

\section{The classical model}
\label{sec:classical_model}
\begin{figure}[!htbp]
       \includegraphics[width=0.4\textwidth]{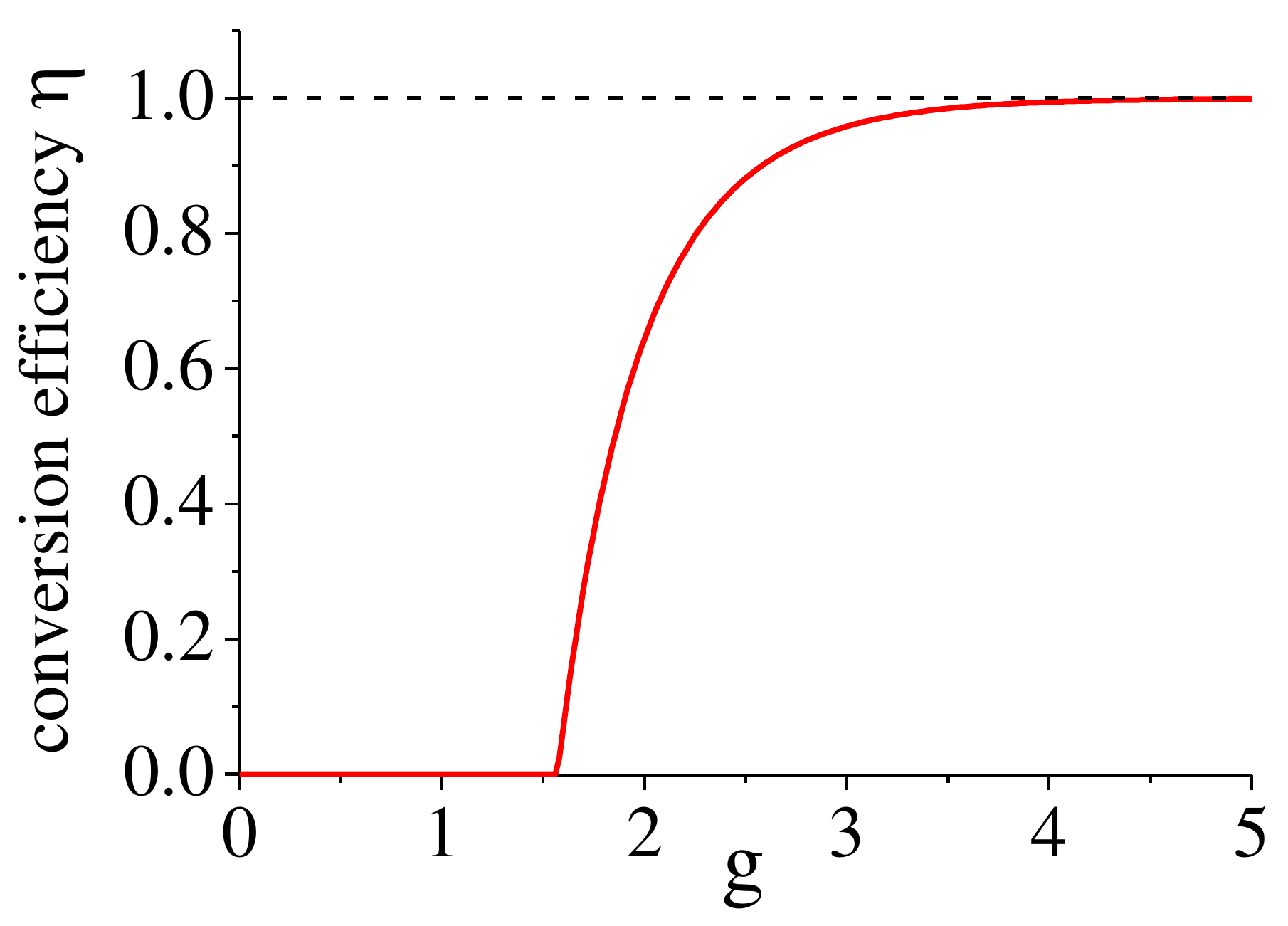}
				\caption{(Color online) Conversion efficiency of the MOPO as a function of the parametric gain $g$ evaluated from Eq.(\ref{eq:ell_int}). The classical model with a CW pump predicts that PDC emission occurs above the MOPO threshold $g=\frac{\pi}{2}$.}\label{fig:classical_model}
\end{figure}

The classical counterpart of the quantum model (\ref{eq:prop_eq}) can be obtained by formally replacing the fields operators $\hat{a}_i$ in Eqs.\,(\ref{eq:prop_eq}) with c-number fields, $\hat{a}_i\rightarrow\alpha_i$, which corresponds to consider $\hat{a}_i=\alpha_i+\delta\hat{a}_i$ and to neglect the quantum fluctuations $\delta\hat{a}_i$. In this way one obtains propagation equations for the c-number fields $\alpha_j(\Omega,z)$ which are formally identical to Eqs.\,(\ref{eq:prop_eq}). In order to recast them in a form which is more familiar in literature, see e.g.\,\cite{Strom,Strom2}, we rather consider the fields:
\bsub
\beqa
\beta_s(z,\Omega)&=&e^{i[k_s(\Omega)-k_s]z}\alpha_s\\
\beta_i(z,\Omega)&=&e^{-i[k_i(\Omega)-k_i]z}\alpha_i\\
\beta_p(z,\Omega)&=&e^{i[k_p(\Omega)-k_p]z}\alpha_p.
\eeqa
\label{eq:beta_fields}
\esub
Then we assume that the effects of second and higher order dispersion are negligible with respect to the first order
\beq
k_j(\Omega)-k_j=k'_j\Omega+\frac{1}{2}k''_j\Omega^2+\hdots\approx\frac{1}{v_{gj}}\Omega,
\eeq
where derivatives $k'_j$, $k''_j$, etc. are calculated at the central frequencies $\omega_j$ and $v_{gj}=1/k'_j$ are the group velocities of the three waves. By back-transforming to the temporal domain $\beta_j(z,t)= \int \frac{d\Omega}{\sqrt{2\pi}} \beta_j(z,\Omega)e^{-i\Omega t}$, the classical propagation equations take the form:
\bsub
\beqa
\frac{\partial\beta_s(z,t)}{\partial z}+\frac{1}{v_{gs}}\frac{\partial \beta_s(z,t)}{\partial t}&=&\chi\beta_p(z,t)\beta_i^*(z,t)\\
\frac{\partial\beta_i(z,t)}{\partial z}-\frac{1}{v_{gi}}\frac{\partial \beta_i(z,t)}{\partial t}&=&-\chi\beta_p(z,t)\beta_s^*(z,t)\\
\frac{\partial\beta_p(z,t)}{\partial z}+\frac{1}{v_{gp}}\frac{\partial \beta_p(z,t)}{\partial t}&=&-\chi\beta_s(z,t)\beta_i(z,t)
\eeqa
\label{eq:classical_prop}
\esub
A non trivial stationary solution to these equations, corresponding to $\partial\beta_j(z,t)/\partial t=0$, is known to exist \cite{Ding}, provided that the injected pump amplitude overcome a suitable threshold value. Assuming that the injected pump is a CW beam, i.e. that $\beta_p(z=0,t)=\bar{\alpha_p}$, and that there is no injected signal and idler wave $\beta_s(z=0,t)=\beta_i(z=0,t)=0$, a non trivial stationary solution, with $\beta_s(z)$ and $\beta_i(z)$ different from zero, exists provided that the dimensionless gain parameter
\beq
g=\sqrt{2\pi}\chi|\bar{\alpha}_p| l_c\,\,\,\,>\,\,\,\,\,\,g_{thr}=\frac{\pi}{2}.
\eeq
Above this threshold, the fraction of down-converted power $\eta=(|\beta_p(z=0)|^2-|\beta_p(z=l_c)|^2)/|\beta_p(z=0)|^2$ (the MOPO conversion efficiency) grows with $g$ as shown by Fig. \ref{fig:classical_model}. More precisely, $\eta$ satisfies the implicit equation
\beq
g=\int_0^{\pi/2}{\frac{d\theta'}{\sqrt{1-\eta\,\sin\theta'}}}\equiv K(\eta),
\label{eq:ell_int}
\eeq
where the transcendental function $K(\eta)$ at the r.h.s is the complete Jacobi elliptic integral of the first kind \cite{Stegun}. It is found (see \cite{Ding})  that this equation has positive solution $\eta>0$ only for $g>\frac{\pi}{2}$. We also notice that that the phases $\phi_s$ and $\phi_i$ of the stationary signal and idler fields above threshold can take arbitrary values, but their sum is always equal to the pump phase $\phi_p$, $\phi_s+\phi_i=\phi_p$.

In this work we are interested in the quantum properties of the PDC field generated from vacuum fluctuations below the threshold (for $g<\frac{\pi}{2}$), where the classical description predicts that the signal and idler waves are identically equal to zero. To this end, we introduce in the next section the linearized model that describes the quantized PDC field  in the undepleted CW pump regime.

\section{Quantum model below threshold}
\label{sec:quantum_model}
Our aim is to investigate the coherence and correlation properties of the down-converted field when approaching the MOPO threshold, describing the transition from a regime of purely spontaneous down-conversion into a regime where most photon pairs are produced through stimulated PDC and the effect of distributed feedback becomes relevant.

In order to perform an analytical treatment, we limit ourself to the case of a monochromatic coherent pump wave of frequency $\omega_p$, assuming its intensity is sufficiently far from the MOPO threshold so that the undepleted pump approximation holds. Under these conditions the pump field can be treated as a known classical field. The corresponding spectral field operator $\hat{a}_p(z,\Omega)$ defined by Eq.(\ref{eq:lin_prop_p}) can be substituted with the $z$-independent c-number function $\alpha_p(\Omega,z=0)=\alpha_p\sqrt{2\pi}\delta(\Omega)$. In this limit, the propagagation equations for the signal and the idler fields take the form:
\bsub
\beqa
&&\frac{\partial}{\partial z}\hat{a}_s(z,\Omega)=+\frac{g}{l_c}\hat{a}_i^{\dagger}(z,-\Omega)e^{-i \mathcal{\bar{D}}(\Omega)z}e^{i\phi_p}\\
&&\frac{\partial}{\partial z}\hat{a}_i^{\dagger}(z,-\Omega)=-\frac{g}{l_c}\hat{a}_s(z,\Omega)e^{i \mathcal{\bar{D}}(\Omega)z}e^{-i\phi_p},
\eeqa
\label{eq:prop_bt}
\esub
where $\phi_p=\text{arg}[\alpha_p]$ is the pump phase at $z=0$ and $\mathcal{\bar{D}}(\Omega)$ is the phase matching function for a monochromatic pump wave, defined by
\beq
\overline{\mathcal{D}}(\Omega)\equiv \mathcal{D}(\Omega,-\Omega)=k_{s}(\Omega)-k_{i}(-\Omega)-k_p+k_G.
\label{eq:CW_pm}
\eeq
The system boundary conditions differ from those found in more common co-propagating geometries. The input field operators, assumed in the vacuum state, are indeed defined at different transverse planes: the left face of the crystal ($z=0$) for the forward propagating signal wave and the right face ($z=l_c$) for the back-propagating idler wave
\bsub
\beqa
\hat{a}_s(\Omega,z=0)&=&\hat{a}_s^{\text{in}}(\Omega)\\
\hat{a}_i(\Omega,z=l_c)&=&\hat{a}_i^{\text{in}}(\Omega).
\eeqa
\esub
By solving Eqs.\,(\ref{eq:prop_bt}) the corresponding output fields $a_s^{\text{out}}(\Omega)=a_s(z=l_c,\Omega)$ and $a_i^{\text{out}}(\Omega)=a_i(z=0,\Omega)$ can be found. The complete field operators [see Eq.\,(\ref{eq:lin_prop})]
\bsub
\beqa
\hat{A}_s^{\text{out}}(\Omega_s)&=&e^{i k_s(\Omega_s)l_c}\hat{a}_s^{\text{out}}(\Omega_s)\\
\hat{A}_i^{\text{out}}(\Omega_i)&=&\hat{a}_i^{\text{out}}(\Omega_i).
\eeqa
\esub
can be expressed in the form of a unitary Bogoliubov transformation, equivalent to the one in \cite{Suhara}:
\bsub
\begin{align}
&\hat{A}_s^{\text{out}}(\Omega)=U_s(\Omega)\hat{A}_s^{\text{in}}(\Omega)+V_s(\Omega)\hat{A}_i^{\text{in} \dagger}(-\Omega)\\
&\hat{A}_i^{\text{out}}(\Omega)=U_i(\Omega)\hat{A}_i^{\text{in}}(\Omega)+V_i(\Omega)\hat{A}_s^{\text{in}\dagger}(-\Omega).
\end{align}
\label{eq:io}
\esub
If we introduce the functions $\phi(\Omega)$, $\beta(\Omega)$, and $\gamma(\Omega)$, defined by:
\beqa
\phi(\Omega)&=&\frac{1}{\cos\gamma(\Omega)-i\frac{\bar{\mathcal{D}}(\Omega)l_c}{2\gamma(\Omega)}\sin\gamma(\Omega)}\\
\beta(\Omega)&=&[k_s(\Omega)+k_i(-\Omega)-(k_s+k_i)]\frac{l_c}{2}\label{eq:phase}\\
\gamma(\Omega)&=&\sqrt{g^2+\frac{\bar{\mathcal{D}}^2(\Omega)l_c^2}{4}},
\eeqa
the gain coefficients $U(\Omega)$ and $V(\Omega)$ can be written as trigonometric functions of the form:
\bsub
\beqa
U_s(\Omega)&=&e^{i k_s l_c}e^{i\beta(\Omega)}\phi(\Omega)\\
V_s(\Omega)&=&e^{i (k_s-k_i) l_c}ge^{i\phi_p}\frac{\sin\gamma(\Omega)}{\gamma(\Omega)}\phi(\Omega)\\
U_i(\Omega)&=&e^{i k_i l_c}e^{i\beta(-\Omega)}\phi^*(-\Omega)\\
V_i(\Omega)&=&ge^{i\phi_p}\frac{\sin\gamma(-\Omega)}{\gamma(-\Omega)}\phi^*(-\Omega)
\eeqa
\label{eq:UV}
\esub
and satisfy the following unitarity conditions
\bsub
\beqa
&&|U_j(\Omega)|^2-|V_j(\Omega)|^2=1,\;\;\;\;\;\;j=s,i\\
&&U_s(\Omega)V_i(-\Omega)=U_i(-\Omega)V_s(\Omega)
\eeqa
\label{eq:unit}
\esub
Notice that $U_j(\Omega)$ and $V_j(\Omega)$ diverge when approaching $g=\pi/2$, the value of the parametric gain corresponding to the MOPO threshold in the stationary CW pump regime \cite{Ding}.
\section{Coherence and correlation}
The quantity of primary interest, which characterizes the twin beams correlation in the spectral domain, is the so-called biphoton correlation:
\beq
\Psi(\Omega_s,\Omega_i)\equiv\langle\hat{A}_s^{\text{out}}(\Omega_s)\hat{A}_i^{\text{out}}(\Omega_i)\rangle.
\label{eq:psi_def}
\eeq
The correlation function (\ref{eq:psi_def}) gives the probability amplitude of finding a signal photon of frequency $\omega_s+\Omega_s$ at the right crystal face and an idler photon of frequency $\omega_i+\Omega_i$ at the left one. Assuming that the signal and the idler input fields are in the vacuum state, and using the input-output relations written in Eq.\,(\ref{eq:io}), we obtain the following expression for the biphoton correlation:
\begin{align}
\Psi(\Omega_s,\Omega_i)&=\delta(\Omega_s+\Omega_i)U_s(\Omega_s)V_i(-\Omega_s)\\
&=\delta(\Omega_s+\Omega_i)e^{i[\phi_p+k_sl_c]}e^{i\beta(\Omega)}\bar{\psi}(\Omega_s),\label{eq:psi_3}
\end{align}
where the $\delta(\Omega_s+\Omega_i)$ function expresses the perfect signal-idler frequency correlation of the monochromatic pump limit. Here we introduced the spectral correlation density
\beqa
\bar{\psi}(\Omega_s)&=&g\; \text{sinc}[\gamma(\Omega_s)]|\phi(\Omega)|^2\\
&=&g\; \text{sinc}[\gamma(\Omega_s)]\left[1+|V_s(\Omega_s)|^2\right].\label{eq:corr1}
\eeqa
The last identity has been obtained from the explicit expression of $U_s(\Omega)$ and $V_i(\Omega)$ given in Eqs.\,(\ref{eq:UV}) and the unitarity condition (\ref{eq:unit}).

Other important quantities are the signal and idler coherence functions
\beq
G^{(1)}_j(\Omega_j,\Omega'_j)=\langle\hat{A}_j^{\dagger\text{out}}(\Omega_j)\hat{A}_j^{\text{out}}(\Omega_j')\rangle\;\;\;\;\;j=s,i
\eeq
which describe the properties of coherence of each of the two fields taken indipendently from the other. From the input-output relations (\ref{eq:io}) it is possible to obtain the following expression for the coherence function:
\begin{align}
\langle A_s^{\dagger \text{out}}(\Omega_s)A_s^{\text{out}}(\Omega'_s)\rangle&=\delta(\Omega_s-\Omega'_s)|V_s(\Omega_s)|^2\\
&=\langle A_i^{\dagger \text{out}}(-\Omega_s)A_i^{\text{out}}(-\Omega'_s)\rangle.
\label{eq:G1}
\end{align}
We wish also to investigate the behaviour of these quantities in the time domain. Precisely, introducing the output temporal fields $\hat{A}_j^{\text{out}}(t)=\int{\frac{d\Omega}{\sqrt{2\pi}}e^{-i\Omega t}\hat{A}_j^{\text{out}}(\Omega)}$, it is possible to write the temporal correlation as:
\begin{align}
\Psi(t_s,t_i)&\equiv\langle\hat{A}_s^{\text{out}}(t_s)\hat{A}_i^{\text{out}}(t_i)\rangle\\
&=g e^{i[\phi_p+k_s l_c]}\int\frac{d\Omega}{2\pi}e^{-i\Omega(t_s-t_i)}\left\{e^{i \beta(\Omega)}\right.\nonumber\\
&\;\;\;\;\;\;\;\;\;\;\;\;\;\;\;\;\;\;\;\;\;\;\;\;\;\;\;\left.\times\text{sinc}[\gamma(\Omega)][1+|V_s(\Omega)|^2]\right\}.\label{eq:biph_spectr}
\end{align}
This function represents the probability amplitude of finding a signal and idler photon at their exit faces at times $t_s,t_i$. The temporal coherence is in turn characterized by
\begin{align}
G_s^{(1)}(t_s,t_s')&\equiv\langle \hat{A}_s^{\dagger\text{out}}(t_s)\hat{A}_s^{\text{out}}(t_s') \rangle\\
&=\int{\frac{d\Omega}{2\pi}e^{i\Omega(t_s-t_s')}|V_s(\Omega)|^2}=G_i^{(1)}(t_s',t_s).
\label{eq:G1_t}
\end{align}
Note that both $\Psi$ and $G^{(1)}$ depend only on the difference $t_s-t'_s$, as it should be for a stationary model.

Approximated analytical expressions of these quantities can be obtained both in the purely spontaneous regime (for $g\ll \frac{\pi}{2}$) and close to the threshold (for $g\rightarrow \frac{\pi}{2}$) by considering the behaviour of the intensity spectrum in these two important limiting cases:
\beqa
&&
S(\Omega)\equiv |V_s(\Omega)|^2=\frac{4g^2\sin^2\gamma(\Omega)}{\bar{\mathcal{D}}^2(\Omega)l_c^2+4g^2\cos^2\gamma(\Omega)}\label{eq:V2c}
\eeqa
\begin{numcases}{\approx}
g^2\text{sinc}^2\frac{\bar{\mathcal{D}}(\Omega)l_c}{2} & for $g\rightarrow 0$ \label{eq:V2b} \\
\frac{4g^2\sin^2 g}{\bar{\mathcal{D}}^2(\Omega)l_c^2+4g^2\cos^2g} & for $g\rightarrow\frac{\pi}{2}$\label{eq:V2a}
\end{numcases}

Further details on the derivation of the spectrum for $g\rightarrow \pi/2$ can be found in Appendix \ref{sec:appendix_b}. 

Performing the expansion of the phase-matching function $\bar{\mathcal{D}}(\Omega)$ (\ref{eq:CW_pm}) and keeping terms up to the first order (phase matching bandwidths in the counterpropagating case are infact extremely narrow) we obtain the approximated relation:
\begin{align}
\frac{ \bar{\mathcal{D}}(\Omega)l_c}{2}&=\frac{l_c}{2}(k_s'+k_i')\Omega+\frac{l_c}{4}(k_s''-k_i'')\Omega^2+\cdots\\
&\approx\frac{\Omega}{\Omega_{gvs}}\label{eq:lin_approx}
\end{align}

where
\beq
\Omega^{-1}_{gvs}\equiv\tau_{gvs}=\frac{1}{2}\left[\frac{l_c}{v_{gs}}+\frac{l_c}{v_{gi}}\right].
\eeq
The inverse ($\tau_{gvs}$) of the characteristic bandwidth $\Omega_{gvs}$ involves the sum of the inverse group velocities rather than their difference: $\tau_{gvs}$ is on the order of the photon transit time across the crystal and represents the typical time delay between counter-propagating twin photon in the spontaneous regime. As we shall see in the next section, in this regime $\Omega_{gvs}$ represents the width of the PDC spectrum.

Another useful approximation needed to perform analytical calculations is the linearization of the phase (\ref{eq:phase}) of the biphoton spectral correlation (\ref{eq:psi_def})
\beq
\beta(\Omega)\simeq(k'_s-k'_i)\frac{l_c}{2}\Omega=\Delta t_A \Omega.
\label{eq:phase_approx}
\eeq
Here 
\beq
|\Delta t_A|=\left|\frac{l_c}{2 v_{gs}}-\frac{l_c}{2 v_{gi}}\right|\ll \tau_{gvs}
\label{eq:lin_approx1}
\eeq
represents the difference of the transit times along the crystal for a pair of counter propagating signal and idler photons generated at the crystal centre at the reference frequencies. 

\section{Low gain regime, $g\ll\frac{\pi}{2}$}\label{sec:low_gain}
We start our analysis from the low gain regime, i.e. $g\ll\frac{\pi}{2}$, where the dominant process is the spontaneous production of photon pairs and distributed feedback does not enter into play.
\subsection{Biphoton correlation}
We consider first the field correlation defined by Eq.\,(\ref{eq:psi_def}) and given by expression (\ref{eq:psi_3}). In the regime of purely spontaneous PDC, $|V_s(\Omega)|^2$ is on the order of $g^2\ll 1$ according to Eq.\,(\ref{eq:V2b}). Its contribution in the expression of the correlation density (\ref{eq:corr1}) is therefore negligible and we have in this limit:
\begin{align}
\lim_{g\rightarrow 0}\bar{\psi}(\Omega_s)&=g\; \text{sinc}\left(\frac{\bar{\mathcal{D}}(\Omega_s)l_c}{2}\right)\label{corr_lg_f1}\\
&\approx g\; \text{sinc}\left(\frac{\Omega_s}{\Omega_{gvs}}\right),&\label{corr_lg_f}
\end{align}
where in the last equality we used the linearized approximation for the phase-matching (\ref{eq:lin_approx}). 

The temporal correlation can be calculated by Fourier transforming the spectral correlation [see Eq.\,(\ref{eq:biph_spectr})]. By using the approximations (\ref{eq:phase_approx}) and (\ref{corr_lg_f}) we recover the result of Suhara for the temporal correlation in  the coincidence count regime \cite{Suhara}. It is given by the box-shaped temporal correlation of width $2\tau_{gvs}$:
\begin{align}
\Psi(\Delta t)&=g e^{i [k_s l_c+\phi_p]}\int{\frac{d\Omega}{2\pi}e^{-i\Omega(\Delta t-\Delta t_A)}\text{sinc}\left(\frac{\Omega}{\Omega_{gvs}}\right)}\label{eq:rect1}\\
&= g\,\frac{e^{i[k_s l_c+\phi_p]}}{2\tau_{gvs}} \text{Rect}\left[\frac{\Delta t-\Delta t_A}{2\tau_{gvs}}\right],\label{eq:rect2}
\end{align}
with $\Delta t=t_s-t_i$ and where we introduced the rectangular function defined as:
	\begin{equation}
	\text{Rect}(x) = \begin{cases} 1 & \mbox{if } |x|<\frac{1}{2} \\ 0 & \mbox{if } |x|>\frac{1}{2}
	\end{cases}
	\label{eq:rect}
	\end{equation}
This function describes a flat distribution of the temporal delays $\Delta t$ between the signal and idler arrival times, ranging between $-\tau_{gvs}+\Delta t_A=-\frac{l_c}{v_{gi}}$ and $\tau_{gvs}+\Delta t_A=\frac{l_c}{v_{gs}}$. As it will be further discussed in Sec.\,\ref{sec:intuitive_picture}, this flat distribution reflects the spontaneous character of the emission in the low gain regime: each photon pair is generated indipendently from the others, and the process can take place at any point of the crystal with uniform probability.

\begin{figure}[!htbp]
   \includegraphics[width=0.45\textwidth]{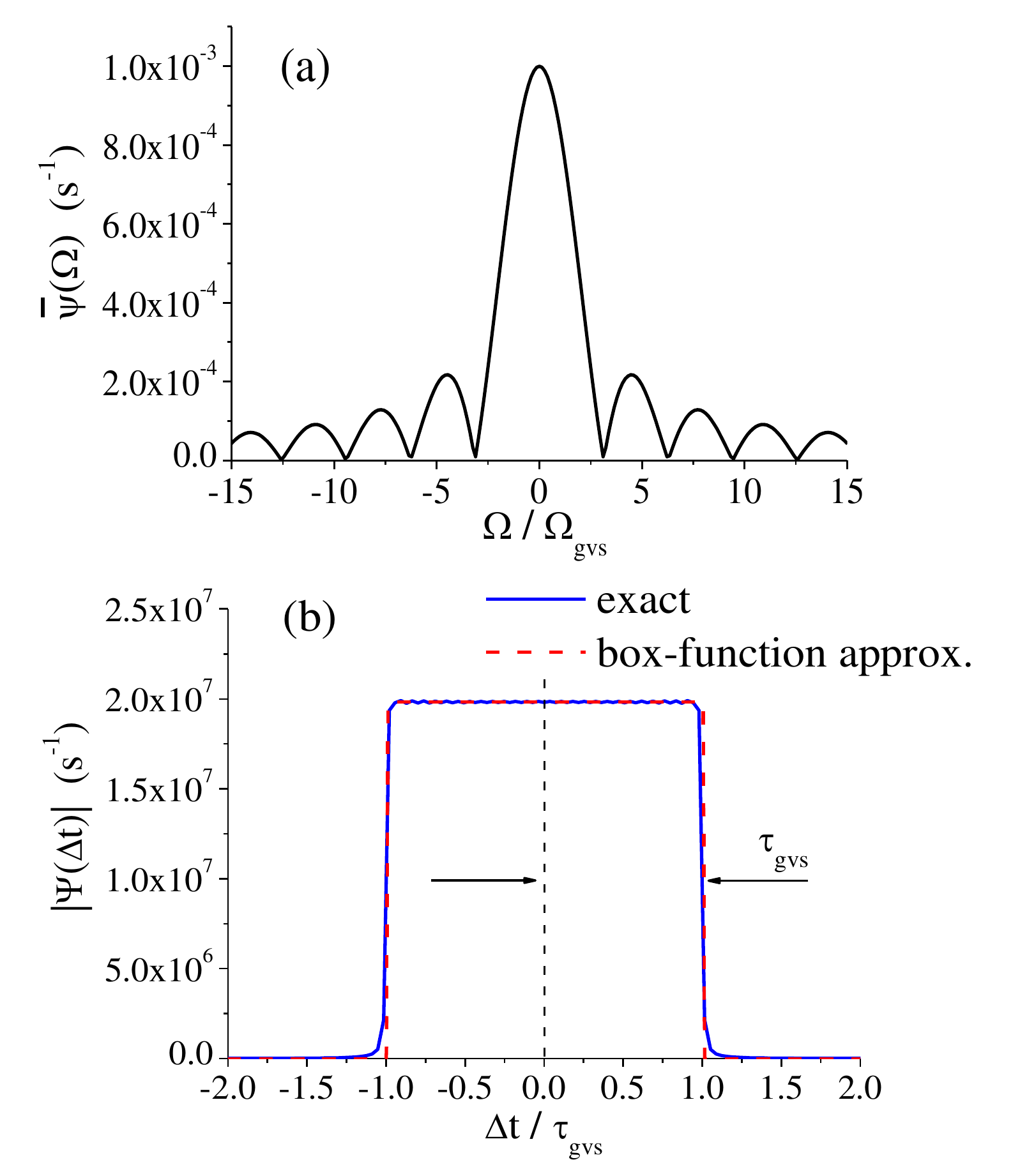}
\caption{(Color online) Biphoton correlation in the spontaneous PDC regime with $g=10^{-3}$ (a) in the spectral and (b) in the temporal domain. In all the figures ''exact'' refers to results obtained from the input-output relations (\ref{eq:io}), without the use of the linear approximations (\ref{eq:lin_approx},\ref{eq:lin_approx1})}.
\label{fig:corr_bt}
\end{figure}

The red curve in Fig.\,\ref{fig:corr_bt}b is the approximate solution (\ref{eq:rect2}), the blue curve is obtained
from the numerical integration of Eq.\,(\ref{eq:biph_spectr}). All the numerical examples reported here and in the following have been obtained for a 4 mm long KTP crystal using the Sellmeier dispersion formula found in \cite{Niko,Kato}. Here we consider the same configuration as in \cite{Can}: Type 0 $e\rightarrow ee$ phase matching for $\lambda_p=821.4$nm, $\lambda_s=1.141$nm, $\lambda_i=2.932$nm. In this configuration $\tau_{gvs}=25.2$ps, $\Delta t_A=-0.55$ps.
\subsection{Coherence function}
In the purely spontaneous regime, the signal and idler spectra [Eq.(\ref{eq:V2c})] are well approximated by Eq.(\ref{eq:V2b}) and (\ref{eq:lin_approx}):
\begin{equation}
	S(\Omega)=|V_s(\Omega)|^2\approx g^2\,\text{sinc}^2\,\left(\frac{\Omega}{\Omega_{gvs}}\right)
	\label{eq:low_gain}
\end{equation}
and exibit the usual squared sinc shape characteristic of the coincidence count regime of PDC.

The coherence function in the time domain is obtained by Fourier transforming the spectrum [Eq.(\ref{eq:low_gain})]. It is characterized by a triangular shape (see Appendix \ref{sec:appendix_a} for a detailed derivation) and can be expressed as a convolution integral over the rectangular biphoton correlation (\ref{eq:rect2}):
  \begin{align}
	G_s^{(1)}(\Delta t)&=g^2\int{\frac{d\Omega}{2\pi}e^{i\Omega \Delta t}\text{sinc}^2\left(\frac{\Omega}{\Omega_{gvs}}\right)}\label{eq:triangular_approx1}\\
	&=\frac{g^2}{2\tau_{gvs}}\,T\left(\frac{\Delta t}{2\tau_{gvs}} \right)\label{eq:triangular_approx}\\
	&=\frac{g^2}{4\tau_{gvs}}\int{dt\,\text{Rect}\left(\frac{t}{2\tau_{gvs}} \right)\text{Rect}\left(\frac{t-\Delta t}{2\tau_{gvs}} \right)}\label{eq:triangular_approx2},
	\end{align}
where the triangle function is
	\begin{equation}
	T(x) = \begin{cases} 1-|x| & \mbox{if } x\in(-1,1) \\0 & \mbox{elsewhere}
	\end{cases}
	\label{eq:triangle}
	\end{equation}
and has the shape of a triangle of base $(-2\tau_{gvs},2\tau_{gvs})$. Therefore the coherence time, taken as the HWHM of the coherence function, is given by half of the sum of the propagation times of the signal and idler photons along the crystal
\beq
\tau_{coh}=\tau_{gvs}.
\eeq
 
\begin{figure}[!htbp]
   \includegraphics[width=0.41\textwidth]{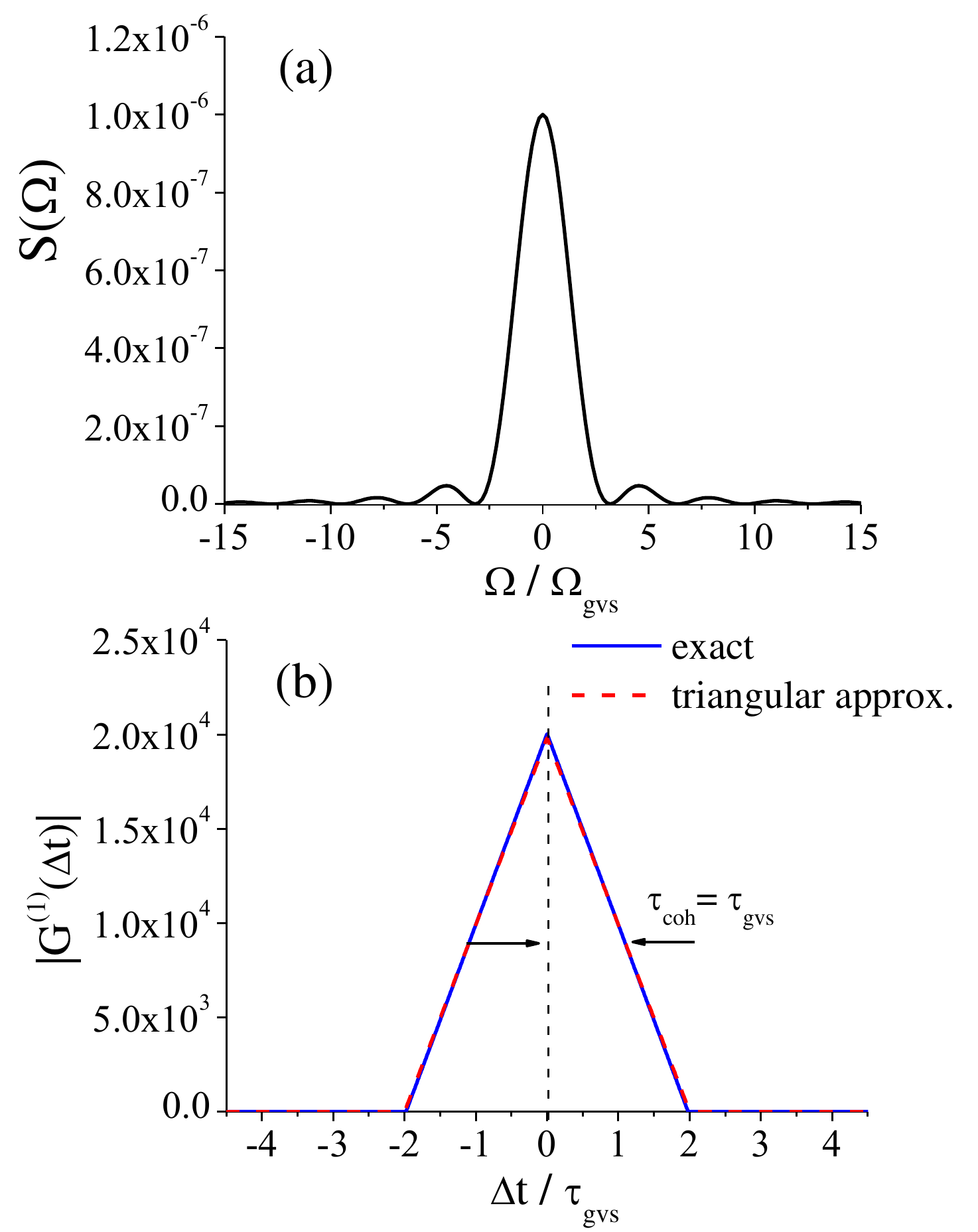}
        \caption{(Color online) (a) PDC spectrum at the crystal output faces in the low gain regime, (b) coherence function in the time domain. In both cases $g=10^{-3}$.}\label{fig:V2g10-3gauss}
\end{figure}
\section{High-gain regime (threshold region), $g\rightarrow\frac{\pi}{2}$}\label{sec:high_gain}
We now consider the regime of stimulated PDC, which occurs when approaching the MOPO threshold from below, i.e. for small positive value of $\epsilon=\frac{\pi}{2}-g$. In this regime, the spectrum is well approximated by the Lorentzian function:
\beq
\lim_{g\rightarrow \pi/2}|V_s(\Omega)|^2=\frac{g^2\sin^2g}{(\Omega^2/\Omega_{gvs}^2)+g^2\cos^2g},
\label{eq:V2_hg}
\eeq
as can be inferred from Eqs.\,(\ref{eq:V2a}) and (\ref{eq:lin_approx}) [see Appendix \ref{sec:appendix_b} for further details]. Such a Lorentzian is characterized by a peak of width (half width at half maximum)
\beq
\Delta\Omega_L=\Omega_{gvs}g\cos g\approx \frac{\pi \epsilon}{2}\Omega_{gvs}\rightarrow 0,\,\,\,\,\,\,\text{for}\,\,\,\,\epsilon\rightarrow 0
\label{eq:fwhm_lor}
\eeq  
which shrinks progressively as the threshold is approached. 
\subsection{Field correlation}
\begin{figure*}[!htbp]
      \includegraphics[width=0.75\textwidth]{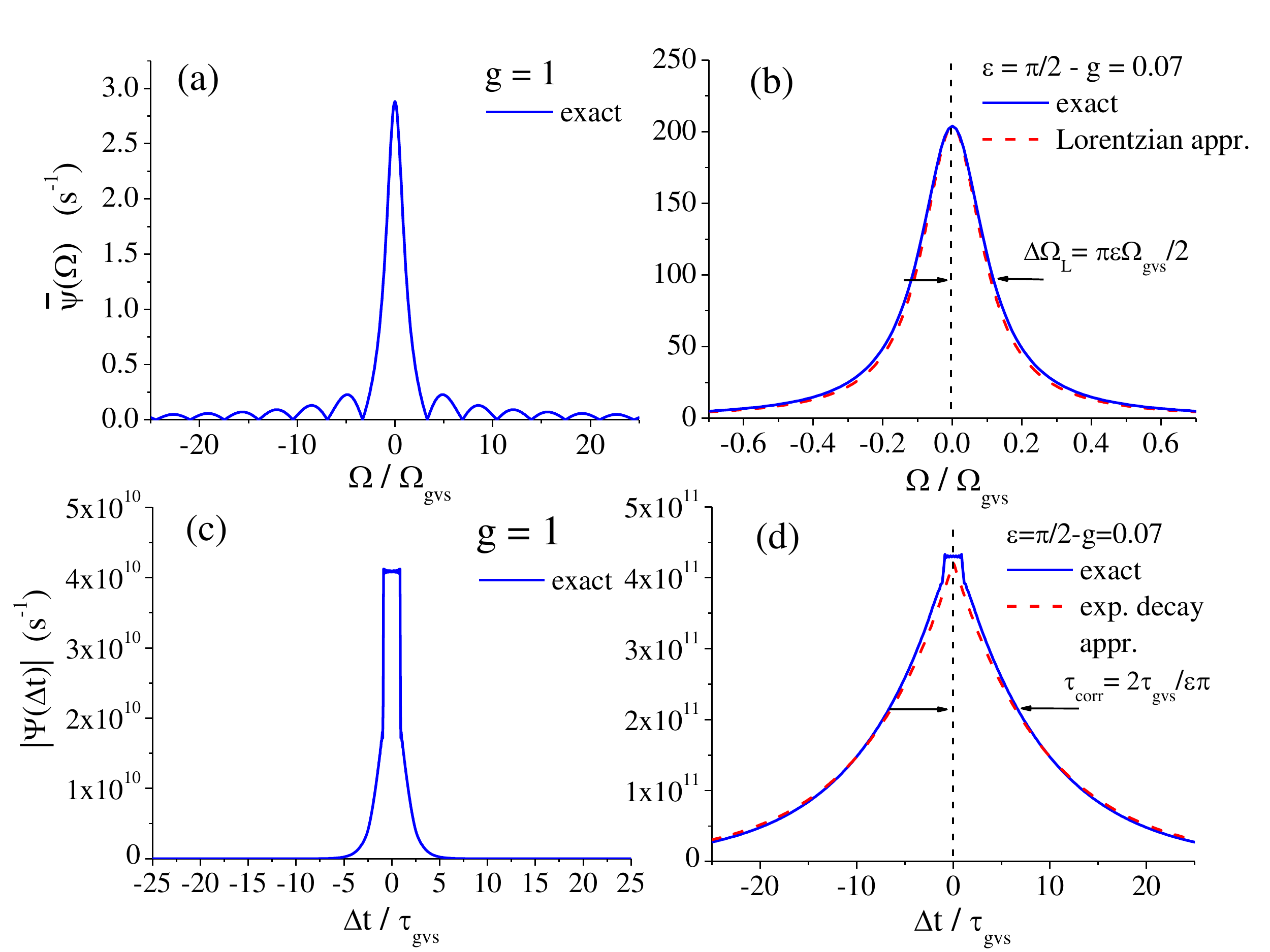}
       				\caption{(Color online) Spectral correlation density (a) at intermediate gain regime $(g=1)$ with both stimulated and spontanous PDC contributing equally. (b) Close to threshold ($\epsilon=\pi/2-g=0.07$) where stimulated PDC is dominant. Biphoton correlation in the temporal domain (c) at intermediate gain and (d) close to thresold.}\label{fig:corr_nt}
\end{figure*}

Based on the Lorentzian approximation (\ref{eq:V2_hg}) for $|V_s(\Omega)|^2$ valid close to threshold, we can write the spectral correlation density  in Eq.\,(\ref{eq:corr1}) as
\begin{align}
\bar{\psi}(\Omega_s)&\approx g\; \text{sinc}[\gamma(\Omega_s)]\left[1+\frac{g^2 \sin^2 g}{(\Omega_s^2/\Omega_{gvs}^2)+g^2 \cos^2 g}\right]\label{eq:biph_f_hg}\\
&\approx g\; \text{sinc}[\gamma(\Omega_s)]+\frac{g^2\sin^3g}{(\Omega_s^2/\Omega_{gvs}^2)+g^2\cos^2g},\label{psi_hg_c}
\end{align}
where in the last line we subsituted $g\; \text{sinc}[\gamma(\Omega_s)]$ in the second term with $\sin g$, since the $\text{sinc}[\gamma(\Omega_s)]$ varies on a scale $\Omega_{gvs}$ which is much broader than the narrow width $\Delta\Omega_L \approx \frac{\pi \epsilon}{2}\Omega_{gvs}$ of the Lorentzian close to threshold. The contribution of stimulated PDC, which increases dramatically close to threshold because of distributed feedback, is responsible of the emergence of this extremely narrow peak [second term in Eq.\,(\ref{psi_hg_c})]. In contrast the smaller contribution [first term in Eq.\,(\ref{psi_hg_c})], similar to the one found in the low gain regime (\ref{corr_lg_f}),  originates from purely spontaneous PDC and extends on a much broader emission bandwidth on the order of $\Omega_{gvs}$. 
Figure \ref{fig:corr_nt}a and \ref{fig:corr_nt}b show the spectral density correlation at an intermediate gain regime and close to threshold respectively. In the latter case the narrow Lorentzian contribution of width $\Delta\Omega_L$ [second term of Eq.\,(\ref{psi_hg_c})] is clearly dominant with respect to the purely spontaneous contribution.
Using approximation (\ref{psi_hg_c}), we find the following expression for the twin beam correlation in the temporal domain:
\begin{align}
\Psi(\Delta t)&\approx g e^{i[k_s l_c+\phi_p]}\left\{\int{\frac{d\Omega}{2\pi}e^{-i\Omega(\Delta t -\Delta t_A)}\text{sinc}\,[\gamma(\Omega)]}\right.\nn\\
&\;\;+\left. \int{\frac{d\Omega}{2\pi}e^{-i\Omega(\Delta t -\Delta t_A)}\frac{g^2 \sin^3 g}{\tau_{gvs}^2\Omega^2+g^2\cos^2g}}\right\},\label{eq:corr_nt_t2}
\end{align}
where $\Delta t=t_s-t_i$. The first term in Eq.\,(\ref{eq:corr_nt_t2}) is on the order of $g/\tau_{gvs}$ and originates from purely spontaneous PDC. The peak value of the second term is $g/(2\tau_{gvs}\cos g)\approx g/(2\tau_{gvs}\epsilon)\rightarrow\infty$ for $\epsilon\rightarrow 0$, and therefore dominates over the first. Thus, close to threshold, we approximately have
\beq
\Psi(\Delta t)\approx\frac{e^{i[k_s l_c+\phi_p]}}{2\tau_{gvs}}\frac{g\, \sin^3 g}{\cos g}e^{-g\cos g \frac{|\Delta t-\Delta t_A|}{\tau_{gvs}}}.\label{eq:corr_nt_t1}
\eeq
The correlation time which characterizes the decaying exponential in Eq.\,(\ref{eq:corr_nt_t1}) 
	\begin{equation}
	\tau_{corr}=\frac{\tau_{gvs}}{g\,\cos g}\approx \frac{2\tau_{gvs}}{\pi \epsilon}\rightarrow\infty\;\;\;\;\text{for}\;\;\;\epsilon\rightarrow 0.
	\label{eq:tau_coh1}
	\end{equation}
goes to infinity for $\epsilon \rightarrow 0$, a feature which reflects the establishment of a feedback effect (see Sec.\,\ref{sec:intuitive_picture}) and which is typical in phase transitions. This behaviour is illustrated in Figs.\,\ref{fig:corr_nt}c - \ref{fig:corr_nt}d  which display the temporal correlaton. In the intermediate regime (Fig.\,\ref{fig:corr_nt}c) where spontaneous and stimulated PDC contribute equally, tails reminiscent of the exponential decay found close to threshold emerge at the basis of the box-shaped correlation characterizing spontaneous PDC. Close to threshold (Fig.\,\ref{fig:corr_nt}d) the size of those tails strongly increases and the correlation is well approximated by the dominant stimulated PDC contribution (\ref{eq:corr_nt_t1}). 
\subsection{Coherence function}
\begin{figure}[!htbp]
      \includegraphics[width=0.4\textwidth]{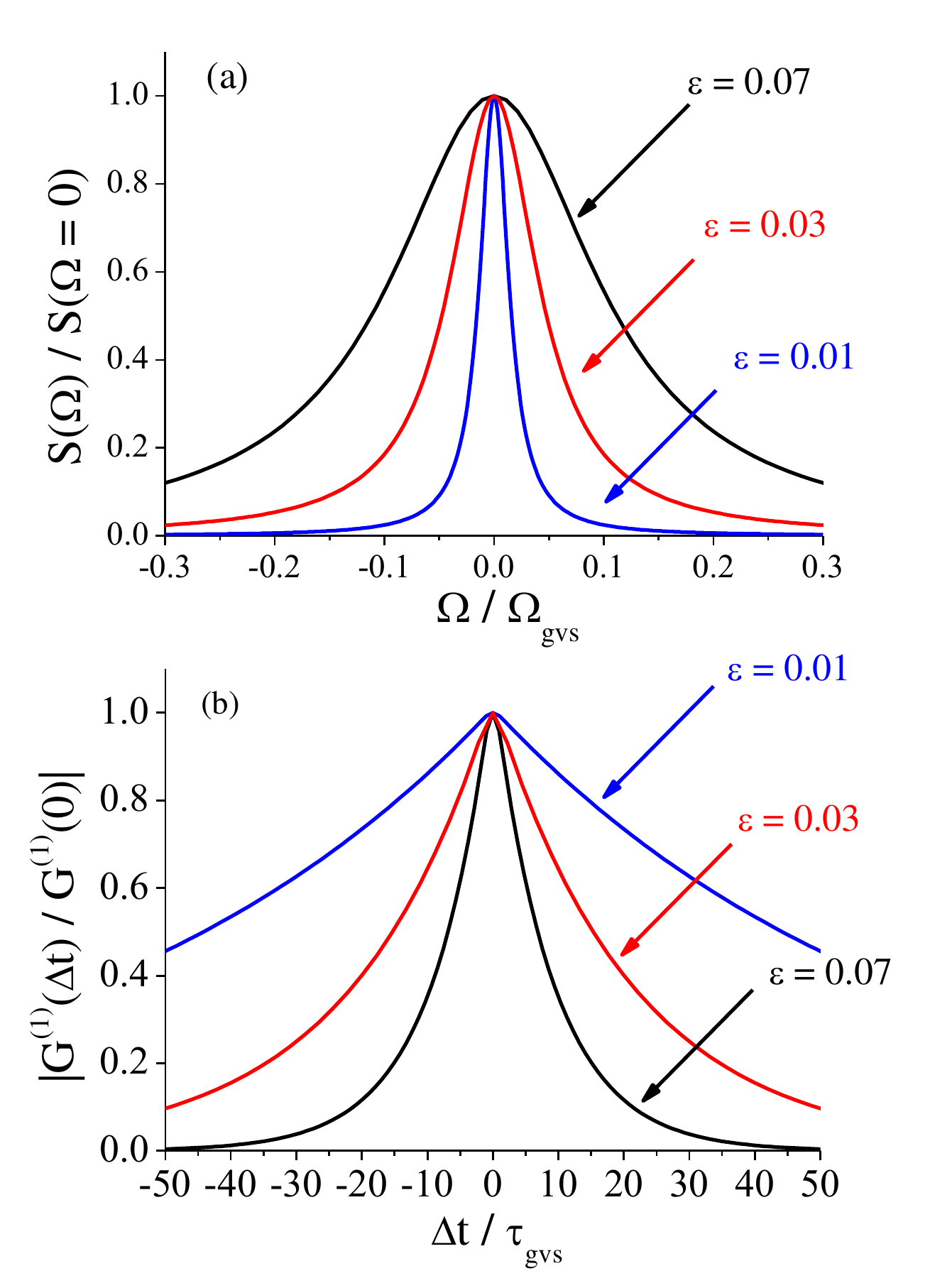}
        \caption{(Color online) (a) Spectrum and (b) temporal coherence function for decreasing values of $\epsilon=\pi/2-g$. The curves in (b) have been obtained through the numerical evaluation of the integrals in Eq.\,(\ref{eq:G1_t}).}\label{fig:appr_thr}
\end{figure}

Close to threshold the PDC emission spectra of the signal and idler fields are well approximated by the Lorentzian function written in Eq.\,(\ref{eq:V2_hg}). The spectrum peak value $|V_s(\Omega=0)|^2=\text{tan}^2 g$ diverges for $g\rightarrow \frac{\pi}{2}$, while its width shrinks to zero for $\epsilon\rightarrow 0$, as for the biphoton correlation, as shown in Fig.\,\ref{fig:appr_thr}a. Clearly, this description based on the linearized model (\ref{eq:prop_bt}) will loose its validity for small but finite values of $\epsilon$, when pump depletion enters into play.

By performing the Fourier transform of the Lorentzian spectrum (\ref{eq:V2_hg}) we obtain the temporal coherence function in the time domain within the same order of approximation:
	\beq
	G_s^{(1)}(\Delta t)=\frac{g}{2\tau_{gvs}}\frac{\sin^2 g}{\cos g}e^{-g\cos g \frac{|\Delta t|}{\tau_{gvs}}}.
	\label{eq:lorentzian_ft}
	\eeq

In contrast to the low gain limit described in Sec.\,\ref{sec:low_gain}, as the MOPO threshold is approached, $G^{(1)}(\Delta t)$ becomes almost indistinguishable from the biphoton correlation (\ref{eq:corr_nt_t1}), apart from the small temporal shift $\Delta t_A$ related to the different group velocities of the signal and the idler fields. Approaching threshold, thus, the coherence and the correlation reflect one the properties of the other because of the cascading processes characteristic of the stimulated regime of pair production. The coherence time which characterizes the decaying exponential in Eq. (\ref{eq:lorentzian_ft}) is the same of the correlation time defined in Eq.\,(\ref{eq:tau_coh1}) and goes to infinity for $\epsilon \rightarrow 0$, i.e.
\beq
\tau_{coh}\approx \frac{\tau_{gvs}}{g\cos g}\approx\frac{2\tau_{gvs}}{\epsilon \pi}\xrightarrow{\epsilon\rightarrow 0}\infty
\eeq
Figure \ref{fig:appr_thr} shows (a) the progressive narrowing of the spectrum and (b) the correspondent broadening of the temporal coherence function for decreasing values of $\epsilon$, a clear manifestation of the critical slowing down of field fluctuations occuring close to threshold.
\section{An intuitive picture}
\label{sec:intuitive_picture}
In this section we want to give an intuitive explanation of the results obtained in sections VI and VII. 

\begin{figure}[!htbp]
  \includegraphics[width=0.45\textwidth]{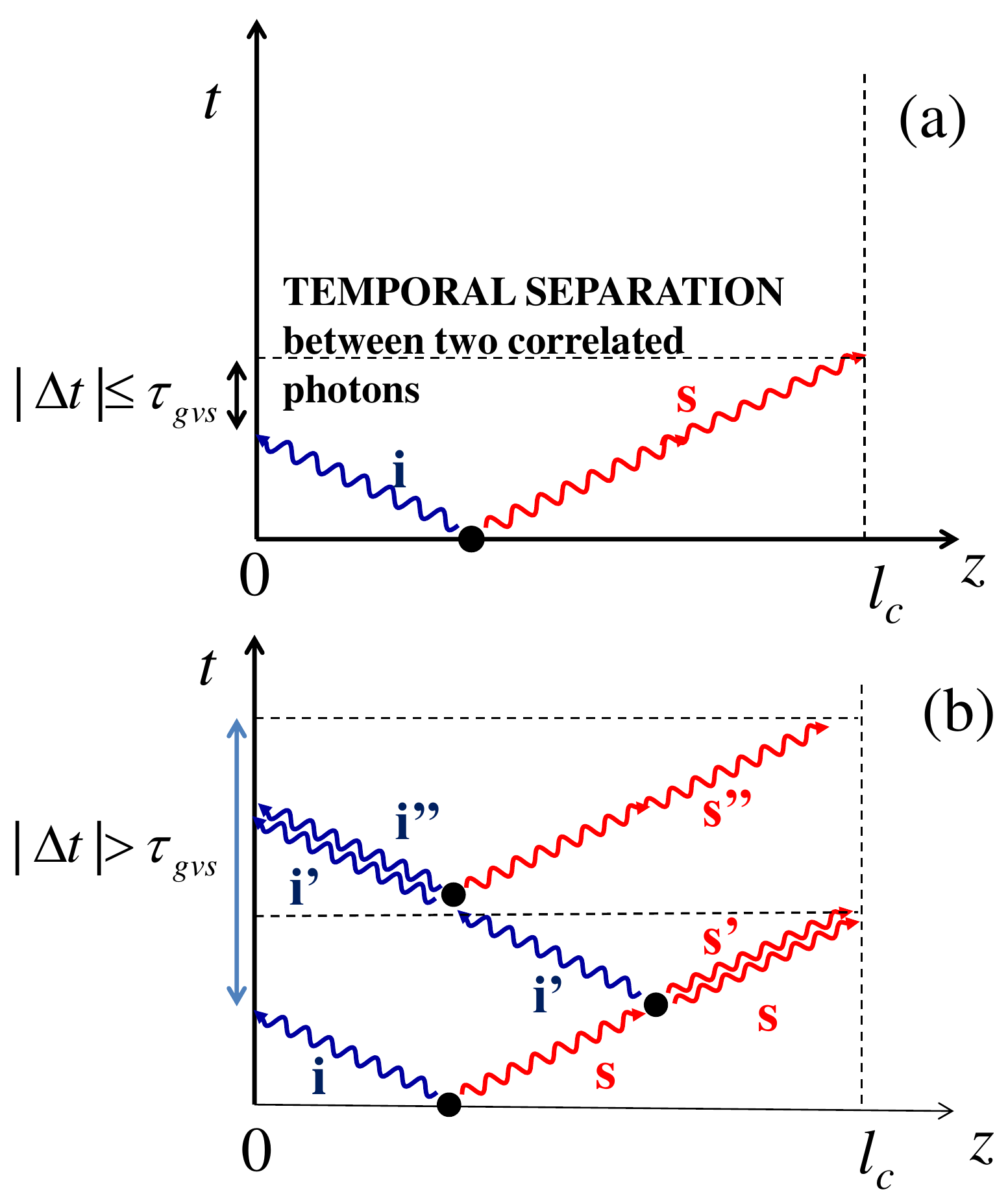}
 \caption{(Color online) Photon pairs originating from a single pump photon at $t=0$ (a) in the purely spontaneous regime and (b) in an intermediate regime where stimulated PDC is not neglible. In case (a) the temporal delay $|\Delta t|$ between the arrival times of two correlated photons cannot be larger than $\tau_{gvs}$, because they originate from the same PDC event. In case (b) $|\Delta t|$ can exceed $\tau_{gvs}$ due to the presence of secondary processes.}\label{fig:pic}
\end{figure}

Figure \ref{fig:pic} represents schematically the propagation of photon pairs originating from a single pump photon produced at time $t=0$ in the $(z,t)$-plane. It considers both a regime of purely spontaneous PDC (Fig.\,\ref{fig:pic}a) and a regime of higher parametric gain where secondary processes due to stimulated PDC take place (Fig.\,\ref{fig:pic}b).

In the first case (Fig.\,\ref{fig:pic}a) the temporal delay $\Delta t=t_s-t_i$ between the arrival times of the twin photons at their respective output faces cannot be larger than $\approx\tau_{gvs}$, as indicated in the figure. If the photon pair is produced close to the crystal center $z=\frac{l_c}{2}$, the two counter-propagating photons exit the crystal almost simultanously (more precisely with a small delay $\Delta t_A=\frac{l_c}{v_{gs}}-\frac{l_c}{v_{gi}}$ due to a possible mismatch of their group velocities). If the pair is produced at $z=l_c$, the signal exits immediately, and the idler arrives at the exit face at $t=l_c/v_{gi}$, thus $\Delta t=-l_c/v_{gi}$. If the pair is produced at $z=0$, conversely, the idler exits immediately while the signal exits the crystal at $t=l_c/v_{gs}$, thus $\Delta t=+l_c/v_{gs}$. The difference of the arrival times is thus strictly within the interval $\left[-\frac{l_c}{v_{gi}},\frac{l_c}{v_{gs}}\right]=\left[\Delta t_A-\tau_{gvs},\Delta t_A+\tau_{gvs}\right]\approx[-\tau_{gvs},\tau_{gvs}]$, since $\Delta t_A\ll\tau_{gvs}$. Well below threshold, where stimulated PDC processes are negligible, each photon pairs is generated indipendently from the others and the probability of generating a photon pair is uniform along the crystal lenght. As a consequence, the distribution of time delays between the two extrema is flat, which explains the box-shaped correlation function displayed in Fig.\,\ref{fig:corr_bt}b.

When the influence of stimulated PDC becomes relevant, the range of allowed values of $\Delta t$ is no more strictly delimited to the interval $[-\frac{l_c}{v_{gi}},\frac{l_c}{v_{gs}}]$.
This is shown in the example of Fig.\,\ref{fig:pic}b where a few secondary processes take place triggered by the first spontaneous twin pair produced at time $t=0$. It is clear from this picture that the arrival times of a pair of signal and idler photons originating from two different elementary processes can differ by a value greater than $\tau_{gvs}$ (see e.g. photon $i$ and $s''$ in the figure). This corresponds to the intermediate gain regime illustrated in Fig.\,\ref{fig:corr_nt}c: purely spontaneous photon pairs and stimulated PDC pairs contribute to the same extent and the biphoton correlation retains its box shaped structure, but with tails developing at the basis that include temporal delays $|\Delta t|>\tau_{gvs}$. 

When approaching threshold (for $g\rightarrow\frac{\pi}{2}$)  most photon pairs are produced through stimulated PDC in a cascading process and much longer temporal delays are allowed. 
In this limit, both the twin photon correlation and the coherence function describing the fluctuations of the individual beams are well approximated by the decaying exponentials (\ref{eq:corr_nt_t1}) and (\ref{eq:lorentzian_ft}) represented in Fig.\,\ref{fig:corr_nt}d and \ref{fig:appr_thr}b. Their characteristic decay times $\tau_{corr}=\tau_{coh}$ becomes much larger than $\tau_{gvs}$ for $g\rightarrow\frac{\pi}{2}$ in agreement with Eq.(\ref{eq:tau_coh1}), as the number of secondary events increases dramatically when approaching the MOPO threshold. 
\section{Conclusions}\label{sec:conclusions}
In this work we provided a theoretical analysis of the coherence and correlation properties of the twin beams generated below the MOPO threshold from a monochromatic CW pump field under stationary conditions. Under the undepleted pump approximation, it was possible to characterize the transition from the regime far from threshold, where the dominant process is the spontaneous production of photon pairs, 
to the regime close to threshold, where the combined effect of stimulated PDC and distributed feedback affects dramatically properties of the light source. 
A narrowing of the spectra and the consequent widening of the correlation and coherence times is predicted when approaching the threshold for coherent emission. This critical slowing down of the quantum fluctuations and critical divergence of the correlation time, which is typical of phase transitions, is studied for the first time in this system. We also gave an intuitive picture 
explaining the main characteristics of the coherence and correlation of the fields in the transition between the low gain and the high gain regime.
\appendix
\section{Coherence function in the low gain regime}
\label{sec:appendix_a}
In this Appendix we evaluate the PDC field coherence function in the low gain limit given by the Fourier transform (\ref{eq:triangular_approx1}) in Sec.V. Using the identity 
\beq
\text{sinc}(u)=\frac{1}{2}\int_{-1}^1{e^{iut}dt},
\eeq 
it can be written as
\begin{align}
	G_s^{(1)}(\Delta t)&=\int{\frac{d\Omega}{2\pi}e^{i\Omega\Delta t}|V(\Omega)|^2}\\
	&=g^2\int{\frac{d\Omega}{2\pi}e^{i\Omega\Delta t}\text{sinc}^2(\tau_{gvs}\Omega)}\\
	&=\frac{g^2}{4}\int_{-1}^1{ds}\int_{-1}^1{ds'}\int{\frac{d\Omega}{2\pi}e^{i[s\tau_{gvs}+s'\tau_{gvs}+\Delta t]\Omega}}
\end{align}
Using the relation $\int_{-\infty}^{\infty}{e^{ius}ds}=2\pi\delta(u)$ and making the substitution $t'=s'\tau_{gvs}$ for the evaluation of the integral in $s'$ we find
\begin{align}
	G_s^{(1)}(\Delta t)&=\frac{g^2}{4\tau_{gvs}}\int_{-1}^1{ds}\int_{-\tau_{gvs}}^{\tau_{gvs}}{dt\,\delta(t+s\tau_{gvs}+\Delta t)}\\
	&=\frac{g^2}{4\tau_{gvs}}\int_{-1}^1{ds\,\text{Rect}\left(\frac{s-\Delta t/\tau_{gvs}}{2} \right)\text{Rect}\left(\frac{s}{2} \right)}\\
	&=\frac{g^2}{2\tau_{gvs}}\,T\left(\frac{\Delta t}{2\tau_{gvs}} \right),
\end{align}
where $T$ is the triangular function defined in Eq.\,(\ref{eq:triangle}).
\section{Lorentzian approximation for the spectrum}
\label{sec:appendix_b}
We provide here a justification of the Lorentzian approximation of the spectrum used for $g \rightarrow \pi/2$ given in Eq.\,(\ref{eq:V2_hg}). We apply the following expansion of $\cos^2\gamma(\Omega)=\cos^2\sqrt{g^2+\left(\frac{\bar{\mathcal{D}}l_c}{2} \right)^2}$ in even power of $\bar{\mathcal{D}} l_c/2$ 
\beq
\cos^2\gamma(\Omega)=\cos^2 g-\frac{\sin g \cos g}{g}\left(\frac{\bar{\mathcal{D}}(\Omega)l_c}{2}\right)^2+\mathcal{O}\left(\frac{\bar{\mathcal{D}}(\Omega)l_c}{2}\right)^4.
\eeq
for evaluating the denominator of the spectrum $|V_s(\Omega)|^2$ given by relation (\ref{eq:V2a}). Keeping only term up to second order we obtain the following approximated expression 
\begin{align}
S(\Omega)&=|V_s(\Omega)|^2\approx\frac{\sin^2\gamma(\Omega)}{\cos^2 g}\frac{1}{1+\frac{2-\sin2g}{2g^2\cos^2g}\left(\frac{\bar{\mathcal{D}}(\Omega)l_c}{2}\right)^2}\\
&\hspace{35mm}\text{for}\,\,\,\,|\bar{\mathcal{D}}(\Omega) l_c|\ll 1\nn
\end{align}
which holds for small value of the phase-mismatch. The key factor lies in that the multiplicative factor of $\bar{\mathcal{D}}^2(\Omega)l_c^2/4$ becomes very large close to threshold, having in this limit
\beq
\frac{2-\sin2g}{2g^2\cos^2g}\approx\frac{4}{\pi^2\epsilon^2}\gg 1\,\,\,\,\,\text{for}\,\,\,\epsilon=\frac{\pi}{2}-g\ll 1.
\eeq
As a consequence, the spectrum $|V_s(\Omega)|^2$ is already reduced by a factor $1/\epsilon^2\gg 1$ with respect to its peak value $\text{tan}^2 g$ as soon as the phase-mismatch becomes on the order of unity, i.e for $|\bar{\mathcal{D}}(\Omega)l_c| \sim 1$. It is thus legimate to use the following approximation
\begin{align}
S(\Omega)=|V(\Omega)|^2&\approx\frac{\sin^2\gamma(\Omega=0)}{\cos^2 g}\frac{1}{1-\frac{2-\sin2g}{2g^2\cos^2 g}\left(\frac{\bar{\mathcal{D}}(\Omega)l_c}{2}\right)^2}\\
&\approx\frac{g^2\sin^2g}{g^2\cos g+\left(\frac{\bar{\mathcal{D}}(\Omega)l_c}{2}\right)^2}
\end{align}
\label{eq:lor_appr}
where in the last approximation we took into account that the multiplicative factor of $(\bar{\mathcal{D}}(\Omega)l_c/2)^2$ is almost equal to unity when $g\rightarrow\pi/2$. Though strictly valid only for frequencies satisfying the condition $|\bar{\mathcal{D}}(\Omega)l_c/2|\ll1$ for which sinc$\gamma(\Omega):=\text{sinc}\sqrt{g^2+(\bar{\mathcal{D}} l_c/2)^2}\approx \text{sinc} g$, 
relation (\ref{eq:lor_appr}) can be extended to the whole frequency domain for the purpose of analytical calculations. According to the previous discussion, $|V_s|^2$  seen as a function of $\bar{\mathcal{D}} l_c/2$ is indeed negligible everywhere except for a narrow neighbourhood of width $\sim \epsilon$ around $\bar{\mathcal{D}} l_c=0$. This neighbourhood translates to a frequency interval on the order of $\epsilon\, \Omega_{gvs}$ when the linear approximation for the phase-matching function (\ref{eq:lin_approx}) is taken into account. 

\bibliography{biblio}
\bibliographystyle{apsrev4-1}
\end{document}